\documentclass[aps,twocolumn,prb]{revtex4}

\usepackage{graphicx}
\usepackage{dcolumn}
\usepackage{bm}
\usepackage{amsmath}
\begin{document}

\newcommand{\bk}{{\bf k}}
\newcommand{\bp}{{\bf p}}
\newcommand{\bpartial}{{\bf \partial}}
\newcommand{\bv}{{\bf v}}
\newcommand{\bV}{{\bf V}}
\newcommand{\bq}{{\bf q}}
\newcommand{\tbq}{\tilde{\bf q}}
\newcommand{\tq}{\tilde{q}}
\newcommand{\bQ}{{\bf Q}}
\newcommand{\br}{{\bf r}}
\newcommand{\bR}{{\bf R}}
\newcommand{\bu}{{\bf u}}
\newcommand{\bb}{{\bf b}}
\newcommand{\bB}{{\bf B}}
\newcommand{\ba}{{\bf a}}
\newcommand{\bs}{{\bf s}}
\newcommand{\bl}{{\bf l}}
\newcommand{\bA}{{\bf A}}
\newcommand{\bK}{{\bf K}}
\newcommand{\bnab}{{\bf \nabla}}
\newcommand{\vd}{{v_\Delta}}
\newcommand{\tr}{{\rm Tr}}
\newcommand{\kslash}{\not\!k}
\newcommand{\qslash}{\not\!q}
\newcommand{\pslash}{\not\!p}
\newcommand{\rslash}{\not\!r}
\newcommand{\bas}{{\bar\sigma}}

\title{Duality and the vibrational modes of a Cooper-pair Wigner crystal}
\author{T. Pereg-Barnea$^1$ and M. Franz$^2$}
\affiliation{$^1$ Department of Physics, The University of Texas at Austin, 
Austin, Texas  78712-1081. \\
$^2$ Department of Physics and Astronomy,
University of British Columbia, Vancouver, BC, Canada V6T 1Z1}
\date{\today}

\begin{abstract}
When quantum fluctuations in the phase of the superconducting order parameter 
destroy the off-diagonal long range order, duality arguments predict
the formation of a Cooper pair crystal. This
effect is thought to be responsible for the static checkerboard patterns
observed recently in various underdoped cuprate superconductors by means of
scanning tunneling spectroscopy. Breaking of the translational symmetry in such a 
Cooper pair Wigner crystal may, under certain conditions, lead to the 
emergence of low lying transverse vibrational 
modes which could then contribute to thermodynamic and transport properties at 
low temperatures. We investigate these vibrational modes using a continuum 
version of the standard vortex-boson duality, calculate the speed of 
sound in the Cooper pair Wigner crystal and deduce the associated
specific heat and thermal conductivity. We then  
suggest that these modes could be responsible for the mysterious 
bosonic contribution to the thermal conductivity recently observed
in strongly underdoped ultraclean single crystals of 
YBa$_2$Cu$_3$O$_{6+x}$ tuned across the superconductor-insulator transition.

\end{abstract}
\maketitle


\section{Introduction}

Ground states of superfluids and superconductors are characterized by a sharply
defined macroscopic phase and an uncertain total number of particles. 
The mathematical
expression of this phenomenon is the number-phase duality: the phase
$\hat{\varphi}$ of the superconducting (or superfluid) order parameter is an 
operator that is quantum mechanically 
conjugate to the particle number operator $\hat{N}$, leading to the uncertainty
relation $\Delta N\cdot\Delta\varphi\geq 1$. A similar duality exists locally.
Consider, for simplicity, a lattice model of a superconductor (or a superfluid)
with the number and phase operators on site $i$ denoted by $\hat{n}_i$ and 
$\hat{\varphi}_i$ respectively. These operators satisfy 
$[\hat{n}_i,\hat{\varphi}_j]=i\delta_{ij}$ which yields the local version
of the above uncertainty relation,\cite{remark0}
\begin{equation}
\Delta n_i\cdot\Delta\varphi_j\geq \delta_{ij}.
\label{uncertain}
\end{equation}
The uncertainty relation (\ref{uncertain}) has an interesting implication
for a state of matter obtained by {\em phase disordering} a superconductor
or a superfluid. (By phase disordering we mean destruction of the 
off-diagonal long range order by phase fluctuations in such a way that the 
{\em amplitude} of the order parameter 
remains finite and large.) In such a 
phase-disordered state the local phase is maximally uncertain, which,
according to (\ref{uncertain}), allows the particle number to be locally sharp.
Cooper pairs (or bosons) may thus minimize their interaction energy by
forming a crystal.

A precise mathematical description of this phenomenon is given in terms of 
vortex-boson duality\cite{dasgupta,lee_fisher,nelson} which maps the system
of Cooper pairs (bosons) near a superfluid-Mott insulator transition onto a 
fictitious dual superconductor in applied external magnetic field. The Cooper
pair (boson) crystal emerges as the Abrikosov vortex lattice of the dual
superconductor. 

Over the past several years it has become increasingly clear that the 
theoretical ideas summarized above may have found a concrete physical 
realization in the underdoped regime of high-$T_c$ cuprate superconductors.
According to one school of thought the ubiquitous pseudogap phenomenon
\cite{timusk} can be thought of as a manifestation of phase-disordered
superconductivity in a $d$-wave channel.\cite{emery1,randeria1,fm1,balents1,ft1} This view is supported by 
a number of experiments\cite{orenstein1,ong1,ong2,sutherland1}
which, in essence, probe the phase fluctuations directly. Given this evidence
and the dual relationship between the number and the phase one is compelled 
to ask whether the expected pair crystallization can be observed in 
underdoped cuprates. The answer to this question appears to be a tentative
``yes'' and we shall elaborate this point below.

The experimental technique of choice to probe for Cooper pair crystallization
is scanning tunneling microscopy (STM). Recent advances in STM provided
a number of atomic resolution measurements of the local density of states 
(LDOS) in a host of cuprate materials. These reveal an intricate
interplay between quasiparticles, vortices, order and disorder 
under the umbrella of high $T_c$ superconductivity.
\cite{Hoffman,McElroy1,Vershinin,McElroy,Hashimoto,Hanaguri}

The STM scans have been performed on large areas of freshly cleaved samples
of Bi$_2$Sr$_2$CaCu$_2$O$_{8+\delta}$ (BiSCCO) and 
Ca$_{2-x}$Na$_x$CuO$_2$Cl$_2$  (Na-CCOC) cuprate superconductors.
Real space maps of LDOS show clear modulations which can be
systematically studied in the Fourier space. Results of the various
measurements can be summarized as follows.  Superconducting samples 
exhibit energy dispersing features at energies that are small compared to the 
maximum superconducting gap. When the samples are underdoped toward the 
pseudogap phase new {\em non-dispersive} features appear. These features 
have periodicity close to four lattice constants but are not 
always commensurate with the underlying ionic lattice.  
At low energies they coexist with the 
dispersing features\cite{McElroy} and are present up to energies of the 
order of the gap maximum, where the dispersing features 
no longer appear.

The appearance of dispersive features in the superconducting phase can
be understood in the context of impurity scattering of the  
quasiparticles.\cite{Wang,Capriotti} The peak-like structures in the Fourier
transformed STM maps  represent the momentum transfer in these scattering 
processes. The latter are sensitive to the quasiparticle dispersion
and the coherence factors\cite{pbf} and thus allow mapping of the underlying
Fermi surface, the gap function, and reflect on the nature of the ordered 
state. The dispersive features also appear in the pseudogap 
phase,\cite{McElroy} indicating that the nodal quasiparticles
survive the transition.  Furthermore, the similarity between 
the observed peaks in the superconducting and pseudogap phases suggests that 
the anomalous (off-diagonal) order parameter is responsible for the pseudogap 
phenomenon.\cite{pbf} The existence of a residual superconducting order 
parameter in 
the insulating phase is consistent with the picture of phase disordered 
superconductivity. 

The non-dispersive features that are seen in underdoped samples indicate the  
formation of additional charge order close to and in the pseudogap phase. 
A crucial characteristic of these modulations is that they leave the low 
energy physics intact. This is manifested by the coexistence of the static 
charge modulations with the low energy dispersing features and by
the V-shaped gap in the density of states. It is therefore difficult to 
imagine that an ordinary charge density wave in the particle-hole 
channel is responsible for the non-dispersive modulations. A conventional
charge density wave would 
gap out the nodal quasiparticles and would therefore change the low 
energy properties of the system, in contrast to 
various experimental results.\cite{Podolsky,franz_sc} 

An interpretation of the static modulations in terms of the Cooper
pair Wigner crystal (PWC) was first suggested by Chen {\it  et al.}\cite{Chen}
who also proposed a test designed to distinguish between PWC and ordinary 
charge density wave.
A detailed theory of the PWC in the context of the phase fluctuation scenario 
was developed by Te\v{s}anovi\'c,\cite{Tesanovic1} who applied the duality
map to the case of a $d$-wave superconductor and demonstrated that 
scrambling 
of the order parameter phase by vortex-antivortex fluctuations indeed 
{\it leads} to Wigner crystallization of Cooper pairs. It was further 
argued that the observed patterns are consistent with a detailed 
account of the phase fluctuating $d$-wave order parameter on the lattice 
with the relevant 
amount of charge carriers in the Cu-O planes.\cite{Melikyan} Using
different variations on this theme, Anderson\cite{anderson1}
and Balents {\it  et al.}\cite{Balents2} arrived at similar conclusions.

In this article we propose and investigate an additional manifestation of 
the existence of a PWC. Cooper pair crystallization should be accompanied 
by emergence of {\em transverse} vibrational modes, absent in the superfluid.
These can be though of as transverse ``magnetophonons'' of the dual 
Abrikosov lattice.
Both the superfluid and the PWC phase support longitudinal modes (phase mode 
or second sound in the former, longitudinal phonon in the latter). For charged 
systems these are gapped due to the long range nature of the Coulomb
interaction and thus cannot be excited at low temperatures. Transverse
modes, on the other hand, correspond to shear deformations of the PWC and 
thus have the
character of gapless acoustic modes. This is strictly true if one can ignore 
the coupling of the PWC to the underlying ionic lattice. If such coupling is 
present then the new modes acquire a gap proportional to the strength
of this coupling. At temperatures smaller than the gap scale such vibrational 
modes would not contribute to thermodynamics or transport.  
We shall assume in the following that this coupling to the lattice is 
negligible although it is not a priori clear why this should be so. The
key hint comes from the fact that at least in some cases the PWC has been
reported to be {\em incommensurate} with the underlying ionic 
lattice.\cite{Vershinin,Hashimoto} This points to very weak coupling, for
if the coupling to the lattice were strong PWC would always be commensurate. 
We shall return to this important point in Sec.\ VIII.

Using the formalism of the vortex-boson duality we calculate the normal modes 
of 
the pair Wigner crystal and note that such modes are accessible through the 
measurement of heat transport. Indeed, Taillefer {\it et al.}\cite{Taillefer1}
have reported an appearance of a new bosonic mode, with  $T^3$ contribution
to the thermal conductivity in YBCO. The onset coincides with
the transition to the pseudogap state at the critical doping $x_c$.
Our calculations below show that this additional 
thermal conductivity may be a result of the PWC lattice vibrations.

This article is organized as follows.  In the next section we briefly review 
a simplified version of the duality transformation that maps a 
phase-fluctuating superconductor onto the ``dual'' superconductor
in an applied
magnetic field. In section III we show that vortices in the dual model 
correspond to Cooper pairs and derive the interaction potential between them.
In section IV we consider an Abrikosov vortex lattice in the dual 
superconductor and calculate its normal modes. In section V we evaluate the 
inertial mass of Cooper pairs (the mass that determines their kinetic 
energy) which is needed to complete the description of the normal modes. 
In the final sections we use the foregoing  results to estimate the 
contribution of the PWC vibrational mode to the specific heat and the
thermal conductivity, summarize our results and discuss their implications
for the physics of cuprates.


\section{Duality and Wigner crystallization of Cooper pairs}

The route to a detailed understanding of the charge modulations in a pair
Wigner crystal goes through a duality transformation. The need for 
the dual description has to do with the fact that  Cooper pairs are highly 
non-local objects, most naturally described in momentum space when they 
form a coherent superconducting state. The duality transformation trades a 
description of a superconductor with fluctuating phase
for a model in which the vortex field plays the primary role. 
The full theory, relevant to cuprates, must take into account the dynamics 
of vortices at the lattice scale in the presence of both Cooper pairs and 
nodal quasiparticles. Its detailed implementation is quite complicated and was 
given in Refs.~[\onlinecite{Tesanovic1,Melikyan}]. Much of the complication 
arises precisely
because of the need to describe phenomena on relatively short length scales,
of the order of the period of the PWC, which is typically close to 4 
lattice spacings. Since we 
shall be concerned primarily with the low-energy, long-wavelength properties
of the PWC, we will contend ourselves with a simpler continuum version of 
the duality transformation\cite{Zee} to which we input the correct lattice 
structure
by hand. As usual, the long-wavelength physics will depend only on the symmetry
and dimensionality of the problem.

We thus consider an effective theory  for superconductivity in a single Cu-O 
plane defined by the partition function $Z=\int{\cal D}[\Psi,\Psi^*]\exp(
-\int_0^{\beta\hbar} d\tau\int d^2 x \tilde{\cal L}/\hbar)$ where $\Psi$ is a
complex scalar order parameter and the Lagrangian density is 
\begin{eqnarray}\label{L0}
\tilde{\cal L} = {1\over 2}\tilde K_\mu\left|\left(\hbar\partial_\mu - 
{2ie \over c}A_\mu\right)\Psi\right|^2  + U(|\Psi|^2).
\end{eqnarray}
The Greek index $\mu=0,1,2$
labels the temporal and spatial components of (2+1) dimensional vectors,
$\partial_\mu=(\partial_{c\tau},\partial_x,\partial_y)$ and $c$ is the speed 
of light in vacuum.
The parameters $\tilde K_\mu=(\tilde K_0,\tilde K_1,\tilde K_1)$ are related to 
the compressibility and phase
stiffness, respectively. They could be calculated, in principle, from an 
underlying microscopic theory via the standard gradient expansion. In the
present work we shall  determine these parameters by matching to the relevant
experimental quantities: $\tilde K_0$ will be expressed through
the Thomas-Fermi screening length and $\tilde K_1$ through the London 
penetration depth.

The electromagnetic vector potential $A$ is explicitly displayed in order
to track the charge content of various fields. In a more general model
one could allow $A$ to fluctuate in space and time and thus implement the 
long range Coulomb
interaction; however we shall not pursue this here. $U$ is a potential
function that sets 
the value of the order parameter $\Psi$ in the superconducting
state in the absence of fluctuations. 

Let us now proceed by disordering the phase of the field $\Psi=
|\Psi|e^{i\theta}$. We fix its amplitude $|\Psi| = \Psi_0$ at the minimum of 
the potential $U(|\Psi|^2)$ and retain the phase $\theta$ as a dynamical 
variable,
\begin{eqnarray}\label{la}
{\cal L} \equiv {\tilde{\cal L}\over\hbar c} = {1 \over 2} K_\mu 
\left(\partial_\mu \theta - {2\pi \over \Phi_0}A_\mu\right)^2 
\end{eqnarray} 
where $K_\mu = (\hbar \tilde K_\mu/ c)\Psi_0^2$ and $\Phi_0 = h c/2 e$ is  
the superconducting flux quantum. To explicitly display the relativistic 
invariance of the above theory it is expedient to rescale 
\begin{equation}\label{res1}
 A_0\to\sqrt{K_1\over K_0} A_0,
\end{equation}
and measure time in natural units related to length as $x_0=c_d\tau$ 
where 
\begin{equation}\label{cd}
c_d=\sqrt{K_1\over K_0}c
\end{equation}
is the speed of the phase mode in the superconductor.
The partition function becomes $Z=\int{\cal D}
[\Psi,\Psi^*]\exp(-\int d^3 x {\cal L})$ with 
\begin{eqnarray}
{\cal L} = {1 \over 2} K 
\left(\partial_\mu \theta - {2\pi \over \Phi_0}A_\mu\right)^2,
\end{eqnarray} 
and $K=\sqrt{K_0K_1}$.

Next we consider fluctuations in the phase. These can be decomposed into 
smooth fluctuations and singular (vortex) fluctuations,
\begin{eqnarray}
\partial_\mu\theta = \partial_\mu\theta_s +\partial_\mu\theta_v \\
\partial\times \partial_\mu\theta_s = 0 
\end{eqnarray} 
The smooth fluctuations are curl-free and the singular fluctuations are related to the density of vortices through their curl in the 
temporal direction, 
\begin{equation}
(\partial\times\partial\theta)_0 = 2 \pi \sum_i q_i \delta(\br - \br_i(\tau)).
\end{equation}
Here $\br_i(\tau)$ are the locations of vortex centers and $q_i$ are the 
respective vortex charges (+1 for vortices and $-1$ for antivortices).

In order to shift our point of view from the condensate to the vortices we 
decouple the quadratic term by an auxiliary field, $W_\mu$, using the
familiar Hubbard-Stratonovich transformation. The resulting Lagrangian is
\begin{eqnarray}\label{eq:W}
{\cal L} = {1 \over 2K}W^2_\mu +iW_\mu(\partial_\mu\theta_v - {2\pi \over \Phi_0}A_\mu)+iW_\mu(\partial_\mu\theta_s)
\end{eqnarray}
We may replace the third term by $-i\partial_\mu W_\mu\theta_s$ (and a 
vanishing surface term) and integrate over the smooth
fluctuations $\theta_s$. This results in the constraint 
$\partial_\mu W_\mu =0$.  In order to enforce this constraint we replace 
the field $W$ by a (2+1) dimensional curl, 
$W_\mu = (\partial \times A_d)_\mu$, and rewrite Eq.~(\ref{eq:W}) as 
\begin{equation}
{\cal L}
={1 \over 2K}(\partial \times A_d)_\mu^2 +i(\partial \times A_d)_\mu\cdot
(\partial_\mu\theta_v - {2\pi \over \Phi_0}A_\mu).
\end{equation}
Integrating by parts in the last term and introducing vortex 3-current
\begin{equation}
j_v = {1\over 2\pi} \partial \times \partial \theta_v
\end{equation}
we obtain
\begin{equation}
{\cal L}=
{1 \over 2K}(\partial \times A_d)_\mu^2 +2\pi i A_d\cdot j_v  -{2\pi i \over \Phi_0}(\partial \times A_d)\cdot A,
\end{equation}
where the dot represents the scalar product in 2+1 dimensions. 
Vortex current $j_v$ is minimally coupled to the gauge field $A_d$.  
This coupling implies that vortices act as electric charges for the dual 
gauge field.

In order to complete our description of the system in terms of vortex
degrees of freedoms 
we introduce a vortex field $\chi$ that is related to the vortex
current through
\begin{equation}
j^\mu_v = i[\chi^* \partial_\mu \chi - (\partial_\mu\chi^*) \chi]
\end{equation} 
We elaborate the theory by adding a kinetic and potential energy terms for the field $\chi$ and write, \cite{Zee,Kleinert}
\begin{eqnarray}\label{eq:dual}
{\cal L} &=& 
{1\over 2}|(\partial_\mu - 2\pi i A_d^\mu)\chi|^2 + {\cal V}(|\chi|)
+{1\over 2K}(\partial \times A_d)_\mu^2
\nonumber \\
&-&{2\pi i \over \Phi_0} A\cdot (\partial \times A_d)+ {\cal L}_{\rm EM}[A], 
\end{eqnarray}
where we have added, for the sake of completeness, the  electromagnetic
Maxwell term for $A$. The inclusion of the potential ${\cal V}$ restores
the physics of short-range 
interactions between the vortex cores that has been neglected when we made
the London approximation in Eq.\ (\ref{la}).

We have now reached our goal of describing a phase fluctuating superconductor
in terms of a vortex field $\chi$ coupled to a dual gauge field $A_d$. If we  
ignore the coupling to electromagnetism, contained in the second line, the
dual Lagrangian formally describes a fictitious superconductor coupled 
to a fluctuating gauge field. As we shall see shortly the temporal component
of the dual magnetic field $B_d=\partial\times A_d$ is related to the charge
density and, thus, the dual superconductor is generally subjected to nonzero
magnetic field. 

At the mean field level this theory exhibits
two distinct phases separated by a second order transition. In a dual 
``normal''  phase, $\langle \chi \rangle = 0$, vortex fluctuations are bounded 
and form finite loops in space-time. This is a phase coherent superconductor
in the direct picture. In the dual ``superconducting'' phase the
vortices condense, $\langle \chi \rangle \neq 0$, and  vortex fluctuations 
proliferate. This is the pseudogap phase in the direct picture where phase 
fluctuations destroy the long range superconducting order. 

Beyond the
mean field  a third possibility exists in which the dual superconductor
has a phase incoherent condensate. In this case $\langle|\chi|\rangle> 0$
but the dual phase coherence is destroyed by quantum melting of the
vortex lattice. This phase is akin to the ``vortex liquid'' phases known 
from the studies of vortex matter.\cite{nelson} A special case is a
situation when most of the dual vortices remain in a crystal but 
some small fraction melts and delocalizes through the sample. This is known 
to happen in a {\em thermally} fluctuating lattice superconductor\cite{teitel}
 when the
magnetic flux per unit cell of the lattice is close to a major fraction,
e.g. $({1\over 2}+{p\over q})$, with $q\gg p$.   In the direct picture 
this situation corresponds to the supersolid phase where the PWC coexists with 
superconductivity. Such coexistence seems to occur in samples of Na-CCOC.
\cite{Hanaguri,franz_sc} Whether such dual supersolid phase occurs in the
class of quantum models considered in this paper is a very interesting open 
question which we shall not attempt to answer here. 


\section{Interaction between Cooper pairs}

The dual model describes a fictitious superconductor. We shall assume that 
this is a type-II superconductor and therefore vortices of the dual field 
can be present.
Vortices in the dual model represent Cooper pairs in the direct picture.  
In order to see this we note that the dual flux is quantized in units of 1:
 compare the term 
$|(\partial_\mu -2\pi i A_d^\mu)\chi|$ in the dual model to 
$|(\partial_\mu -2\pi i\Phi_0^{-1} A_\mu)\Psi|$ in the direct model.
Next let us obtain the relation between the {\it real} electromagnetic charge 
and the dual magnetic field.  The coupling
${2\pi i \Phi_0^{-1}}A\cdot(\partial\times A_d)$ implies that the 
electromagnetic current is given by 
$j_\mu^e/\hbar c = 2\pi \Phi_0^{-1} (\partial\times A_d)_\mu$, or
\begin{equation}
j_\mu^e = 2e (\partial \times A_d)_\mu.
\end{equation}
Specifically, the flux in the temporal direction is related to the charge 
density, $\rho = 2e (\partial \times A_d)_0$.
This means that a vortex in $\chi$ carries the electric charge
\begin{eqnarray}
Q_v = \int \rho(\br) d^2r = 2e \int (\partial \times A_d)_0 d^2 r= 2e,
\end{eqnarray}
where the integration is over a surface containing the vortex.

\begin{figure}
\includegraphics[width=7cm]{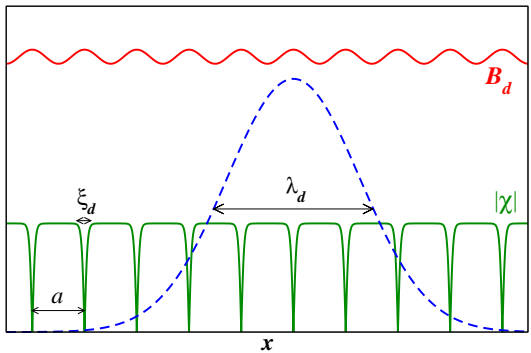}
\caption{(Color online) A schematic plot of the spatial variation of the order parameter 
$\chi$ and the field $B_d$ in the dual Abrikosov vortex lattice. $\xi_d$ and 
$\lambda_d$ denote the dual coherence length and dual penetration depth, 
respectively. Dashed line represents the magnetic field $b_d(\br)$
associated with a
single isolated vortex; $B_d$ in a dense lattice is a superposition of such 
field profiles centered at individual vortices. }
\label{fig1}
\end{figure}

With nonzero density of Cooper pairs the dual superconductor is subject to
dual magnetic field, $B_{d0}=(\partial\times A_d)_0$. If we assume that 
we are in the dual type-II limit, a dual 
Abrikosov vortex lattice is formed.
From now on we shall  drop the adjective `dual' whenever 
there is no potential for confusion. In the vortex lattice the magnetic field 
exhibits a periodic modulation with one flux quantum per unit cell.
In the direct picture this corresponds to a {\em charge density modulation} 
with charge  $2e$ per unit cell. This is the pair Wigner crystal. Figure
\ref{fig1} illustrates the spatial variation of the order parameter $\chi$ and 
the field $B_d$ in such a dual Abrikosov lattice.

We are interested in the interaction 
between vortices which we regard as point particles located at the phase
singularities associated with each vortex. We consider a static configuration
of vortices located at points $\br_i$ and neglect fluctuations in both
$\chi$ and $A_d$. This corresponds to a dual mean field approximation. We 
emphasize that this is a highly nontrivial mean field theory since it
describes the original superconductor in the presence of strong quantum
fluctuations. From Eq.\ (\ref{eq:dual}) the energy of a collection of static 
vortices becomes
\begin{equation}
E_{\rm MF}={\hbar c\over 2}\int d^2r\left[|(\nabla-2\pi i\bA_d)\chi|^2
+K_0^{-1}(\nabla\times\bA_d)^2\right],
\label{emf1}
\end{equation} 
where, for simplicity of notation, we regard $\bA_d$ as a 3-dimensional vector
with $z$ component always zero. We have also passed back to conventional units
reversing the scaling introduced in connection with Eq.(\ref{res1}).

The interaction energy is easiest to evaluate
in the dual London approximation, where we assume a constant amplitude of the
order parameter $|\chi(\br)|=\chi_0$, but arbitrary variation in its phase 
$\alpha$. As in the case of ordinary superconductors the London approximation
is valid when the intervortex distance $a\gg\xi_d$ the dual coherence length;
i.e.\ the vortex cores do not overlap. $\xi_d$ is difficult to estimate 
reliably
since it depends on ${\cal V}$ in Eq.\ (\ref{eq:dual}) which has been 
essentially added by hand to account for the short-distance behavior of 
(real) vortices. We shall assume that $\xi_d$ is of the order of ionic lattice
spacing deep in the dual superconducting phase. Close to the transition
it diverges (as $\sim \chi_0^{-1}$ in the mean field theory) and thus we 
expect the London approximation to break down in this limit.

To find the minimum of the energy we regard $E_{\rm MF}$ as a functional of 
$\bA_d$ and $\alpha$ and vary it with respect to the gauge field 
\begin{equation}
{\delta E_{\rm MF} \over \delta A_d} = 0,
\end{equation}
with result
\begin{equation}\label{eq:maxwell}
2\pi\chi_0^2(\nabla \alpha - 2\pi \bA_d) = K_0^{-1} \nabla \times \bB_d.
\end{equation}
Now we can rewrite the energy as
\begin{equation}
E_{\rm MF}={\hbar c \over 2K_0} \int d^2r [\lambda_d^2(\nabla \times \bB_d)^2 
+ \bB_d^2]
\label{emf2}
\end{equation}
where
\begin{equation}\label{eq:lD}
\lambda_d^2 = {1 \over 4 \pi^2 \chi_0^2 K_0}.
\end{equation} 
$\lambda_d$ is the dual penetration depth which in the context of the Abrikosov
lattice characterizes the magnetic size of a vortex. In the direct picture
it represents the size of the charge cloud
associated with a single Cooper pair.

If we apply the curl operation to Eq.\ (\ref{eq:maxwell}) we obtain the dual
London equation 
\begin{equation}\label{london}
-\lambda_d^2\nabla^2\bB_d+\bB_d=\hat{z}n(\br)
\end{equation} 
where 
\begin{equation}\label{density}
n(\br)=\sum_i\delta(\br-\br_i)
\end{equation} 
is the vortex density. The delta functions  appear 
due to multiple valuedness of the phase in the presence of vortices; 
$\nabla\times\nabla\alpha(\br) =2\pi\hat{z}\sum_i\delta(\br-\br_i)$.
It is clear that $\bB_d=\hat{z}B_d$. The solution of  the London equation
(\ref{london}) can be written as   
\begin{equation}\label{bb}
B_d(\br) = \int d^2r' G(\br-\br')n(\br'),
\end{equation} 
where $G(\br)$ is a Green's function subject to 
\begin{equation}\label{green}
(1-\lambda_d^2\nabla^2)G(\br)=\delta(\br).
\end{equation} 
The above equation has a simple solution in the Fourier space,
\begin{equation}\label{green_f}
G(\bk)={1\over 1+\lambda_d^2k^2}.
\end{equation} 

Using a vector identity $(\nabla\times\bB_d)^2=\bB_d\cdot(\nabla\times\nabla
\times\bB_d)+\nabla\cdot(\bB_d\times\nabla\times\bB_d)$ and discarding
the vanishing surface term we rewrite the vortex energy (\ref{emf2}) as  
\begin{equation}
E_{\rm MF}={\hbar c \over 2K_0} \int d^2r \bB_d\cdot
[-\lambda_d^2\nabla^2\bB_d+\bB_d].
\label{emf3}
\end{equation}
With the help of  Eqs.\ (\ref{london}) and (\ref{bb}), this can be recast as a
density-density interaction
\begin{equation}
E_{\rm MF}={1\over 2} \int d^2r\int d^2r' n(\br)V(\br-\br')n(\br'),
\label{emf4}
\end{equation}
where we have introduced the intervortex potential 
$V(\br)=(\hbar c/K_0)G(\br)$. In view of Eq.\ (\ref{density}) one can also
write this as interaction between point-like particles located at $\br_i$,
\begin{equation}
E_{\rm MF}={1\over 2} \sum_{ij}V(\br_i-\br_j).
\label{emf5}
\end{equation}

For future reference it is useful to give the intervortex potential in terms
of physically accessible quantities. Specifically, we can trade compressibility
$K_0/\hbar c$ for the Thomas-Fermi screening length $\lambda_{TF}$. In 
conventional terms the latter is given by\cite{am} 
\begin{equation}
{1 \over \lambda_{TF}^2} = 4\pi e^2 {1\over d}
{\partial n_0 \over \partial \mu} 
\end{equation} 
where $n_0$ is the two dimensional density of particles 
and we have added the factor of $1/d$ to account for the layer thickness. 
We thus have $\hbar c/K_0=4\pi e^2\lambda_{TF}^2/d$. This allows us to write 
an explicit expression for the intervortex potential in the Fourier space
as 
\begin{equation}\label{eq:int}
V(\bk) = {e_r^2  \over \bk^2+\lambda_d^{-2}},
\end{equation} 
with the effective charge
\begin{equation}\label{er}
e_r^2 = {4\pi e^2\over d}  \left({\lambda_{TF} \over \lambda_d}\right)^2.
\end{equation}

Two remarks are in order. First, despite its appearance $V(\br)$ is {\em not} 
electrostatic in origin; $e^2$ appears simply because we chose to express
compressibility in terms of the Thomas-Fermi screening length. The interaction
is mediated by phase fluctuations in the physical superconductor. Second,
it is useful to consider the interaction in real space. One obtains 
$V(\br)=(e_r^2/2\pi)k_0(r/\lambda_d)$ where $k_0(x)$ is the Hankel function 
of 0-th order. Of interest is its asymptotic behavior,
\begin{equation}\label{eq:int2}
V(\br) = {e_r^2  \over 2\pi}\Biggl\{
\begin{array}{ll}
\left({\pi\over 2}{r\over\lambda_d}\right)^{1/2}e^{-r/\lambda_d},\ \ & r\gg\lambda_d \\
\ln\left({r\over\lambda_d}\right) +0.12, & r\ll \lambda_d \\
\end{array}
\end{equation} 
Thus the interaction has a finite range, of the order of the dual penetration
depth. Approaching the transition, $\lambda_d$ diverges.
The long distance behavior of $V$ indicates that in this limit the interaction
becomes long ranged but, at the same time, its strength vanishes since 
$e_r\to 0$.


\section{Normal modes in a Cooper pair Wigner crystal}\label{section:modes}

The problem of normal modes in a lattice with long range Coulomb interaction 
has been studied previously in the context of electronic 
Wigner crystals.\cite{EWigner}  In two dimensions, vibrations of a square
lattice with centrally symmetric interactions between electrons contain
modes with imaginary frequencies indicating that the lattice is,
in general, unstable. In the dual picture this corresponds to the well known
fact that while the triangular Abrikosov vortex lattice is stable, the 
square lattice
represents an energy maximum and is, therefore, unstable.\cite{kleiner}
It is also known that the square vortex lattice can be stabilized if the
interactions posses four-fold anisotropy.\cite{berlinsky1} Such
anisotropies can arise from the $d$-wave symmetry of the order parameter
\cite{affleck} or from band structure effects associated with the underlying
ionic lattice.\cite{kogan1} In the following we shall make an assumption that
similar terms with four-fold anisotropy exist in our dual superconductor. 
These terms will stabilize the square vortex lattice but will not affect
our discussion of the long wavelength vibrational modes.

We employ the standard formalism for lattice vibrations: we assume that
vortices are located at points $\br_i=\bR_i+\bu_i$ where $\bR_i$ denotes
vectors of a Bravais lattice and $\bu_i$ are small displacements.
The interaction potential (\ref{emf5}) is expanded to
second order in the displacements,
\begin{equation}
V(\bR_{ij}+\bu_{ij}) \simeq V(\bR_{ij}) + u_{ij}^\alpha 
D^{\alpha\beta}_{ij}u_{ij}^\beta,
\end{equation}
where $\bR_{ij} = \bR_i - \bR_j$, $\bu_{ij}=\bu_i-\bu_j$ and
\begin{eqnarray}\label{eq:tildeV}
D_{ij}^{\alpha\beta} &=&
{\partial^2 V(\bR_{ij}+\bu)\over\partial u^\alpha\partial u^\beta}
\bigg|_{\bu=0}\nonumber \\
&=& -e_r^2\int {d^2k \over (2\pi)^2}{k_\alpha k_\beta \over \bk^2 + \lambda_d^{-2}}e^{i \bk \cdot\bR_{ij}}
\end{eqnarray}
is the dynamical matrix.
Aside from a constant $V_0 =  \sum_{ij}V(\bR_{ij})$, the elastic energy
can be written as 
\begin{equation}
\Delta E =  2e_r^2(I_1 -I_2)
\end{equation}
with
\begin{eqnarray}
I_1&=& \sum_iu_i^\alpha u_i^\beta\sum_j D^{\alpha\beta}_{ij}, \\
I_2 &=& \sum_{ij}u_i^\alpha u_j^\beta D^{\alpha\beta}_{ij}.
\end{eqnarray}

Note that the intervortex potential $V$ is defined everywhere in space
(not only on the lattice 
sites) and therefore $\bk$ is not restricted to the first 
Brillouin zone.  In order to keep all momentum integration variables within 
the Brillouin zone we replace $\bk \to \bq+\bQ$, and  
\begin{equation}
\int{d^2 k} \to \sum_{\bQ}\int_{\rm BZ} {d^2 q} \\
\end{equation}
where $\bQ$ is a reciprocal lattice vector ($e^{\bQ\cdot\bR}=1$ for any 
lattice vector $\bR$).

We now analyze the terms $I_1$ and $I_2$ by Fourier transforming the 
variables $\bu_i$,
\begin{equation}\label{eq:fourier}
\bu_i = a^2\int_{\rm BZ} {d^2k \over (2\pi)^2}e^{i \bk\cdot\bR_i}
\bu_{\bk}.
\end{equation}
The momentum integral is over the first Brillouin zone and $a$ is the vortex 
lattice constant which we include in order for $\bu_\bk$ to maintain the
dimension of length.
Combining equations (\ref{eq:tildeV}-\ref{eq:fourier}) we find
\begin{eqnarray}
I_n &=& -\int(d\bk)(d\bk')(d\bq)u_\bk^\alpha u_{\bk'}^\beta \sum_{\bQ}{(q_\alpha+ Q_\alpha)(q_\beta+ Q_\beta) 
\over (\bq + \bQ)^2 + \lambda_d^{-2}} \nonumber \\
&\times& a^4\sum_{ij}  \begin{cases} e^{-i(\bk+\bk')\cdot\bR_i+\bq\cdot(\bR_i-\bR_j)} & n=1\\
	 e^{-i\bk\cdot\bR_i-i\bk'\cdot\bR_j+\bq\cdot(\bR_i-\bR_j)} & n=2
	\end{cases}
\end{eqnarray}
where $(d\bk) = d^2k / (2\pi)^2$ and the integration extends over the first 
Brillouin zone. Performing the real space sums leads to
\begin{equation}\label{dele}
 \Delta E = 2 e_r^2\int(d\bk)u_\bk^\alpha u_{-\bk}^\beta Z_{\alpha\beta}(\bk),
\end{equation}
with
\begin{equation}\label{ze}
Z_{\alpha\beta}(\bk)=\sum_{\bQ}\left[{(k_\alpha+ Q_\alpha)(k_\beta+ Q_\beta) 
\over (\bk + \bQ)^2 + \lambda_d^{-2}}
-{ Q_\alpha Q_\beta \over Q^2 + \lambda_d^{-2}}\right].
\end{equation}

The normal modes of the system are related to the eigenvalues $z_i(\bk)$ of the 
matrix  $Z_{\alpha\beta}(\bk)$.  Explicitly $m_v\omega_i^2 = 2 e_r^2 z_i(\bk)$ 
where $m_v$ is the dual vortex mass. We thus need to evaluate the reciprocal
lattice sums indicated in Eq.\ (\ref{ze}). These sums are slowly convergent
and great care must be taken in evaluating them; specifically we need to 
employ the Ewald summation technique. The following derivation is similar to that 
given by Fetter\cite{Fetter} which was done in the context of a triangular 
Abrikosov vortex lattice.

In order to proceed analytically we split $Z_{\alpha\beta}(\bk)=
Z^{(1)}_{\alpha\beta}(\bk)+Z^{(2)}_{\alpha\beta}(\bk)$ with 
\begin{equation}\label{z1}
Z^{(1)}_{\alpha\beta}(\bk) = {k_\alpha k_\beta \over \bk^2+\lambda_d^{-2}} 
\end{equation}
and $Z^{(2)}_{\alpha\beta}(\bk)$ containing the sum over all $\bQ\neq 0$.
Evaluation of the latter is greatly simplified if we assume that 
$\lambda_d^{-1}$ is much smaller than the smallest reciprocal lattice vector, 
$Q_0\simeq 2\pi/a$. Our estimate of  $\lambda_d$ below will justify this 
assumption. Thus,  
\begin{equation}\label{z2}
Z_{\alpha\beta}^{(2)}(\bk) \approx \sum_{\bQ\neq 0}\left[
{(k_\alpha+ Q_\alpha)(k_\beta+ Q_\beta) 
\over (\bk + \bQ)^2}-{ Q_\alpha Q_\beta \over \bQ^2}\right].
\end{equation}
We note that all the dependence on $\lambda_d$ is contained in 
$Z^{(1)}_{\alpha\beta}(\bk)$ which is, in addition, independent of the lattice
structure. $Z^{(2)}_{\alpha\beta}(\bk)$, on the other hand, depends on the 
lattice structure but is independent of $\lambda_d$.

The detailed calculation of $Z^{(2)}_{\alpha\beta}(\bk)$ is given in the appendix.
The result can be summarized as  
\begin{equation}\label{zab}
Z_{\alpha\beta}(\bk) = {k_\alpha k_\beta \over k^2 + 
\lambda_d^{-2}}+{\vartheta \over 2}a^2(\delta_{\alpha\beta} k^2 - 2k_\alpha k_\beta)
\end{equation}
where for the square lattice $\vartheta= 0.066$. Eq.\ (\ref{zab}) is valid 
for long wavelengths, $k\ll 2\pi/a$. 

\begin{figure}
\includegraphics[angle = -90,width=\columnwidth]{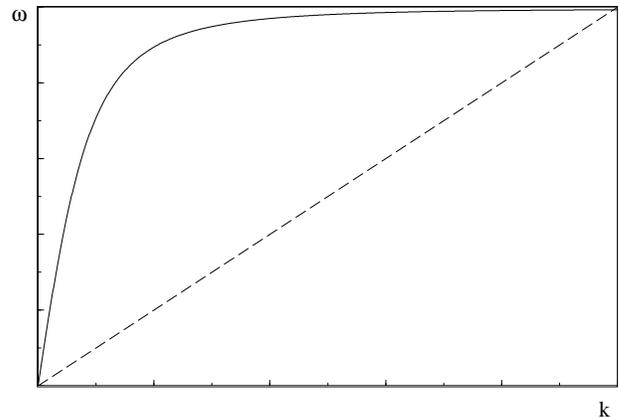}
\caption{A schematic plot of the acoustic transverse mode (dashed) and the 
longitudinal mode (solid).  The functions are plotted with 
$\lambda_d = 10 a$ where $a$ is the vortex lattice constant and $\omega(k)$ is presented in units of $v_s$.}\label{fig:dispersion}.
\end{figure}

The normal modes are readily found 
through the eigenvalues of the matrix $Z_{\alpha\beta}(\bk)$,
\begin{eqnarray}
\omega_1(\bk) &=& v_s k, \\
\omega_2(\bk) &=& v_s k \sqrt{{2\over \vartheta a^2}{ 1 \over k^2 + \lambda_d^{-2}} - 1}, 
\end{eqnarray}
where 
\begin{equation}\label{vs}
v_s^2 = \vartheta {e_r^2\over m_v}.
\end{equation}
The first, acoustic mode, is transverse and its sound velocity is $v_s$.  
The second mode is longitudinal. At long wavelengths, $k\ll\lambda_d$,
the longitudinal mode is also acoustic with velocity 
$v_s(\lambda_d/a)\sqrt{2/\vartheta}$. 
As the intervortex interaction becomes long-ranged ($\lambda_d\to\infty$)
the latter speed diverges and the mode becomes gapped, which is expected on
general grounds. 
With the estimated $\lambda_d$ of about 10 vortex lattice constants the 
longitudinal mode is unimportant (as we shall see it is the inverse of 
the sound velocity that enters the expressions for the specific heat and thermal 
conductivity). Also, had we retained the long-range Coulomb interaction
mediated by the electromagnetic gauge field fluctuations, the longitudinal
mode would be gapped for any value of $\lambda_d$.
A schematic plot of the two modes is given in Fig. \ref{fig:dispersion}.

%


\section{Dual vortex mass}\label{section:mass}

In order to evaluate the sound velocity of the  transverse 
acoustic mode it is essential to estimate the mass of the dual vortex.  
Naively this mass is simply the Cooper pair mass, $2m_e$.  However, this is not the parameter $m_v$ that enters the sound 
velocity expression.  The difference between the two originates from our scheme of calculating the normal modes.
We have modeled the PWC as a system of point masses and springs. This 
analogy is useful but not quite suitable for the case of Cooper
pairs that are delocalized over many lattice sites.  The finite size of the 
Cooper pairs 
causes significant overlap between their wave functions and this leads to a 
situation that is quite different from the familiar case of vibrations
of point-like ions in a solid. 

To make these considerations more concrete let us now estimate
the effective Cooper pair size  given by the dual penetration 
depth $\lambda_d$. 
According to the STM experiments the charge modulation amounts to $\nu=1-5$\% 
of the total charge density.\cite{Davis} In a dual vortex lattice this 
corresponds to the rms amplitude variation of the dual magnetic field 
$\nu\approx\langle(B_d-\bar{B}_d)^2\rangle^{1/2}/\bar{B}_d$, where 
$\bar{B}_d$ is the 
average field and the angular brackets denote the spatial average. 
The above rms variation depends on the ratio of $\lambda_d$ to
intervortex spacing $a$.\cite{amin} Namely, $a^2/\lambda_d^2\approx \nu$,
which leads to the estimate
\begin{equation}\label{lambda_d}
\lambda_d\approx (5-10)a.
\end{equation}
Near half filling we can estimate $a\approx \sqrt{2}a_0\approx 5$\AA~,
where the ionic lattice constant $a_0=3.8$\AA~ in YBCO.
Thus, even deep in the dual superconducting phase, the charge and mass of
the Cooper pair are distributed over many lattice sites.

In the present case it is intuitively clear that only the fraction of the
Cooper pair mass associated with the pair density wave should contribute
to the vibrational degrees of freedom. In particular as $\lambda_d\to\infty$
the charge distribution becomes homogeneous, the system is superfluid and
cannot support any transverse modes. We thus seek a mass parameter associated
with the kinetic energy of a moving dual vortex. The appropriate parameter 
is the {\em inertial mass} of a dual vortex. In order to determine the latter
we note that, if we discard the coupling to electromagnetic field, the 
vortex dynamics are given by a relativistic theory Eq.\ (\ref{eq:dual}).
Correspondingly, the rest energy of a dual vortex is given by 
\begin{equation}\label{rest}
E_{\rm rest} = m_v c_d^2.  
\end{equation}
Here, $m_v$ is the inertial mass of the dual vortex that we seek and $c_d$ is 
the dual speed of light, i.e., the phase velocity of the dual gauge field
defined in Eq.\ (\ref{cd}).
We may therefore estimate the energy of a vortex line $E_{\rm rest}$ 
in the standard way\cite{Tinkham} and deduce its inertial mass through Eq.\
(\ref{rest}).

We consider a single vortex located at the origin. Its energy can be expressed
by combining Eq.\ (\ref{emf3}) with the dual London equation (\ref{london})
and $n(\br)=\delta(\br)$ as
\begin{equation}
E_{\rm rest} = {\hbar c \over 2 K_0}b_d(0).
\end{equation}
Henceforth we shall denote magnetic field associated with a {\em single} 
isolated vortex by $b_d(\br)$.
The energy of a vortex 
depends linearly on the magnetic field at its center.
For a circularly symmetric vortex the London equation can be solved by the 
Hankel function
\begin{equation}
b_d(r) = {1 \over 2\pi \lambda_d^2 }k_0\left({r\over \lambda_d}\right)\approx {1 \over 2\pi \lambda_d^2 }
\left[\ln\left( {\lambda_d \over r }\right) + 0.12\right]
\end{equation} 
where the last equality holds for $r\ll\lambda_d$. The divergence as 
$r\to 0$ is unphysical and occurs due to our neglect of the order parameter 
amplitude variation in the vortex core. A reliable estimate is obtained 
by evaluating 
\begin{equation}
b_d(0)\approx b_d(\xi_d) \approx {1 \over 2\pi \lambda_d^2 }\ln(\kappa_d)
\end{equation} 
where $\kappa_d = \lambda_d /\xi_d$ is the dual Ginzburg-Landau parameter.  
$\kappa_d$ is assumed larger than unity 
(dual type-II regime) but since it appears inside the logarithm its exact 
value is unimportant.

The dual vortex mass is thus given by
\begin{equation}\label{mv}
m_v = {\hbar \over 4 \pi c \lambda_d^2 K_1} \ln(\kappa_d) = {4 e^2 \over d c^2} \left({\lambda \over \lambda_d} \right)^2 \ln (\kappa_d).
\end{equation}
We have used the  London penetration depth, 
\begin{equation}\label{lam}
\lambda^{-2} = {4\pi e^2 n_s\over m_ec^2}
\end{equation}
with the superfluid density $n_s=2\Psi_0^2/d$ to eliminate $K_1 = \hbar \Psi_0^2/2m_e c$. 
In the above estimate one should take
the mean field value of $\lambda$, i.e.\ the value it would have in the absence
of phase fluctuations. In YBCO, we thus take $\lambda\simeq 1000$\AA,
the value at optimum doping. Taking $\kappa_d=10$ and $d=12$\AA~  gives
$m_v\simeq 1.1 \times 10^{-5}(\lambda/\lambda_d)^2(2m_e)$, where $m_e$ is the 
electron mass. 
	
A more instructive way of expressing $m_v$ is to estimate the mean field
superfluid density 
close to half filling by $n_s\approx 1/(2a_0^2d)$ and obtain
\begin{equation}
m_v = 2m_e\left({a_0\over \lambda_d}\right)^2{\ln (\kappa_d)\over 2\pi}.
\end{equation}
As expected, when Cooper pairs are localized and $\lambda_d$ approaches 
the lattice constant $a_0$ the inertial mass of the
dual vortex approaches that of the  Cooper pair. When, on the other hand,  
$\lambda_d\gg a_0$, the Cooper pair is delocalized over many lattice
spacings and the dual vortex mass becomes small.


\section{Interlayer tunneling and dual monopoles}

A linear dispersion such as the one we found for the transverse mode, 
combined with Bose-Einstein distribution for phonons, leads to specific
heat $C_v\sim T^d$ where $d$ is the dimensionality of the system. The 
thermal conductivity, within the simple Boltzmann approach, is 
$\kappa = {1 \over 3} C_v v_s \ell$ where $\ell$ denotes the phonon mean free
path. Assuming the latter is $T$-independent, as is the case for phonons
scattered by the sample boundaries, we have $\kappa\sim T^d$. We thus arrive 
at a conclusion that, in order to agree with experiment, the PWC phonons
must propagate in all 3 dimensions. 

Our theory thus far focused on purely two dimensional physics of the 
Cu-O layers. However, it is clear that vibrations
of PWC in the adjacent layers will be coupled. There are two main sources 
of this coupling: (i) the pair tunneling between the layers represented
by the interlayer Josephson term and (ii) the Coulomb interaction between
the induced charge modulations. Inclusion of the interlayer coupling will 
cause the PWC phonons to propagate between the layers. To complete our
calculation we must determine the associated sound velocity in the $\hat{z}$
direction. 

Ideally, we would like to extend our duality map to a system of weakly 
coupled layers and repeat the calculation of the normal modes. Unfortunately, 
there is no straightforward generalization
of the vortex-boson duality to a system in 3+1 dimensions; 2+1 dimensional
systems are special in this respect. The reason for this can be seen most 
clearly by returning back to Eq. (\ref{eq:W}). In (3+1)D it is not possible
to enforce the zero-divergence constraint on $W_\mu$ by expressing it as a curl
of a gauge field; the curl operation is only meaningful in 3 dimensions. 
On a more 
fundamental level in  (3+1)D vortices cannot be regarded as point 
particles, rather they should be thought of as strings. Thus, rather than a 
dual superconductor, in (3+1)D the duality map produces a {\em string theory}.
 On
physical grounds we still expect the PWC to form in a (3+1)D phase
disordered superconductor but the underlying mathematical structure of
the dual theory appears to be more complicated and beyond the scope
of this paper.\cite{remark1,strings}

We thus continue describing the physics of the individual layers by the (2+1)D
duality and consider the effect of weak interlayer coupling on the resulting 
quantum state. We focus first on the Josephson coupling, which mediates 
tunneling of Cooper pairs between layers. Formally we imagine generalizing 
our starting Lagrangian (\ref{L0}) into a Lawrence-Doniach 
model\cite{Tinkham,doniach1} for a multilayer system by attaching a layer
index $m$ to the scalar field $\Psi$ and adding a term 
\begin{equation}\label{don}
{J_0\over 2 a_0^2}|\Psi_m-\Psi_{m+1}|^2
\end{equation}
to couple the layers. The cross terms describe tunneling of physical
Cooper pairs between the layers. The amplitude for this 
process, $J_0$, is related to the $c$-axis London penetration depth $\lambda_c$
by 
\begin{equation}\label{lamc}
J_0={a^2_0 \over d}{\hbar^2c^2 \over 16 \pi e^2\lambda_c^2}.
\end{equation}

\begin{figure}
\includegraphics[width=8.4cm]{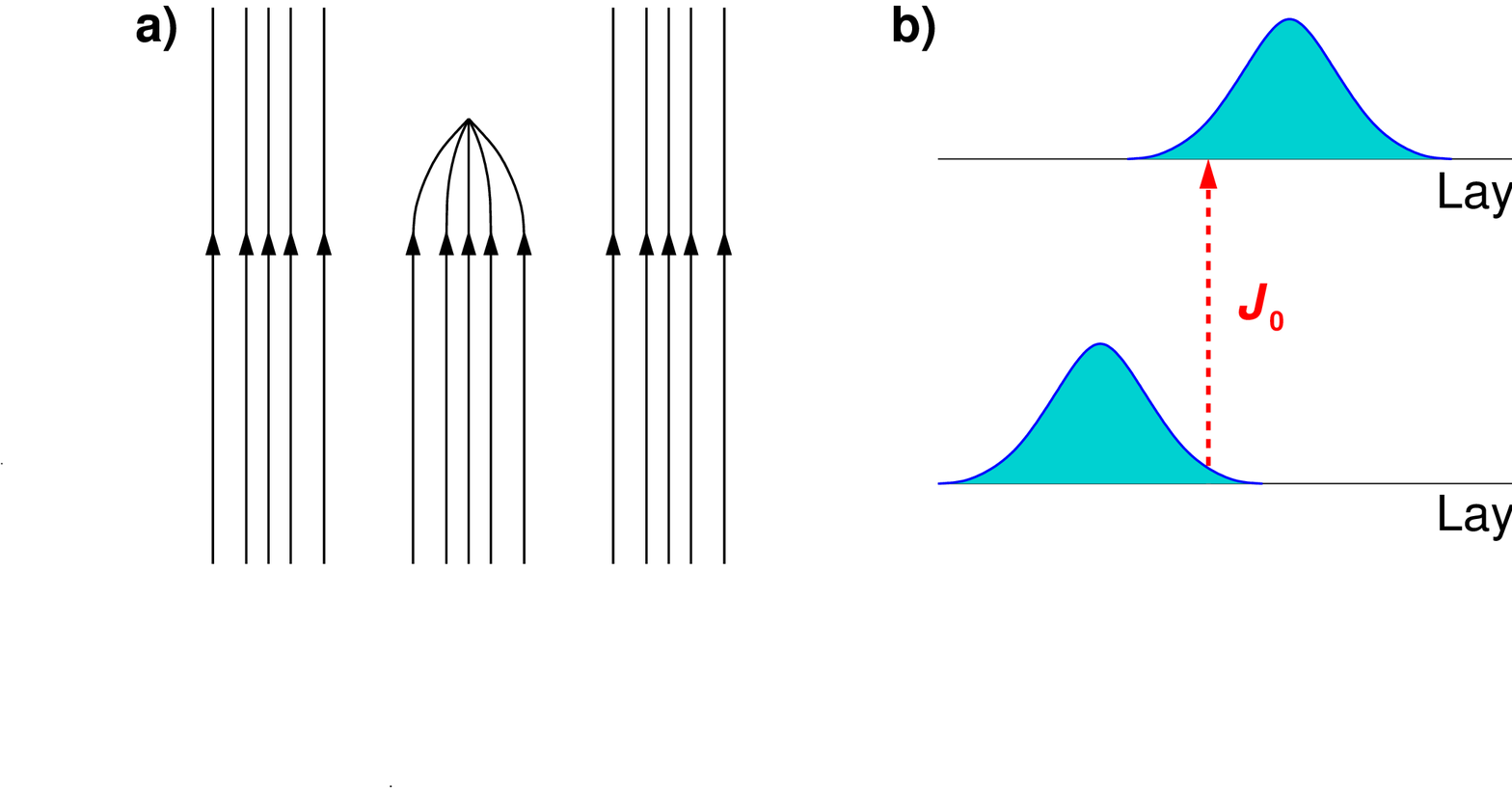}
\caption{(Color online) a) A schematic illustration of the flux lines in the presence of a
monopole. b) Josephson tunneling with lateral displacement of the dual
vortex. The shaded regions represent charge densities associated with a 
Cooper pair in the PWC. }
\label{fig3}.
\end{figure}

From the point of view of an  
individual layer removal of a Cooper pair represents a {\em monopole event}.
A tunneling event occurring at a particular instant of imaginary time 
adds or removes a Cooper pair from the plane. In the dual description,
this event corresponds to the appearance or disappearance of a vortex; 
the magnetic flux lines associated with it originate or
terminate at the same point, as illustrated in Fig.\ref{fig3}a.   
A point in space-time which represents a drain (source) of a magnetic flux is 
known as magnetic monopole (antimonopole). 
The vortex must reappear in the adjacent layer and this
corresponds to an antimonopole. In ordinary superconductors vorticity is 
strictly
conserved which can be regarded as a consequence of the absence of monopoles
in the real world. In the dual superconductor vorticity is conserved only 
if we consider a system with fixed number of Cooper pairs. Once we introduce
the Josephson tunneling vorticity is conserved only globally (i.e. the total 
number of vortices in all layers is constant) and we must permit 
monopole-antimonopole pairs to occur in the adjacent layers.

To model the interlayer tunneling we should thus regard the dual gauge
field $A_d$ as {\em compact} and append to the dual Lagrangian (\ref{eq:dual}) 
a term describing monopole-antimonopole events occurring in the adjacent 
layers. Since, ultimately, we only
need the mean field description of the intervortex interaction, we shall adopt
a simpler and physically more intuitive approach. In essence, all we need
to complete our description of PWC vibrations is an effective potential,
akin to $V(\br_i-\br_j)$, that will describe the coupling between vortices in 
adjacent layers induced by the Josephson tunneling. Taking our clue from
Eq.\ (\ref{emf5}) we describe the system with many layers by a Hamiltonian
$H=H_0+H_J$ with
\begin{eqnarray}
H_0 &=&{1\over 2}\sum_m\int_\br\int_{\br'} \psi^\dagger_m(\br) \psi^\dagger_m(\br')
V(\br-\br')\psi_m(\br') \psi_m(\br), \nonumber \\
H_J &=&\sum_m\int_\br\int_{\br'}J(\br-\br')[\psi^\dagger_m(\br)\psi_{m+1}(\br')
+{\rm h.c.}],
\end{eqnarray}
The  operator $\psi^\dagger_m(\br)$ creates a vortex at point $\br$ in the 
$m$-th layer and $\int_\br=\int d^2r$.
 $H_0$ describes interactions between vortices within a layer
and $H_J$ generates the Josephson coupling; $J$ is the amplitude for 
interlayer tunneling to be discussed shortly. For now we regard vortices
as infinitely heavy (no kinetic energy in the plane). 

Our strategy will
be to treat $H_J$ as a small perturbation on the eigenstates of $H_0$. This
is justified as long as $J$ is very small. In the
limit of infinite mass the unperturbed eigenstates are labeled by 
the positions $\{\br_{im}\}$
of $N$ individual vortices within each layer. The energy of this 
state is simply given by 
Eq.\ (\ref{emf5}). Clearly, there will be no correction to the energy to first 
order
in $H_J$. The leading contribution appears in second order and
describes a virtual hop of a Cooper pair from one layer to the next and then
back. For simplicity we consider a case of two layers, $m=1,2$, and evaluate 
the second order correction to the energy from such processes:
\begin{equation}\label{e2}
\Delta E^{(2)}=\sum_{\eta=\pm 1}\int_\bs\int_{\bs'}
{|\langle N,N|H_J|N+\eta_\bs,N-\eta_{\bs'}\rangle|^2
\over E(N,N)-E(N+\eta_\bs,N-\eta_{\bs'})}.
\end{equation}
In an obvious notation $|N,N\rangle$ denotes a state with $N$ pairs in each 
layer, located at $\{\br_{i1}\}$ and $\{\br_{i2}\}$, while 
$|N+\eta_\bs,N-\eta_{\bs'}\rangle$ represents a state with one pair added
(removed) at $\bs$ in layer 1, and one pair removed (added) at $\bs'$ in layer
2, for $\eta=1 (-1)$.

The matrix element in the numerator is easily seen to be just 
$|J(\bs-\bs')|^2$ but the energy denominator requires some careful thought.
An obvious contribution to the energy difference comes from the interaction
between Cooper pairs expressed through Eq.\ (\ref{emf5}). For $\eta=+1$
this is simply the interaction energy of a vortex added at $\bs$ in layer 1
and antivortex added at $\bs'$ in layer 2,
\begin{equation}\label{den1}
\Delta U=-\sum_i[V(\bs-\br_{i1})-V(\bs'-\br_{i2})].
\end{equation}
For $\eta=-1$ the overall sign is reversed. However, this cannot be the 
whole story
since Eq.\ (\ref{den1}) does not account for the fact that the state with 
$N$ vortices per layer
corresponds to the absolute minimum of the total energy of the unperturbed 
system. At the level of 
Eq.\ (\ref{emf5}) this requirement is implemented as a constraint. This 
is adequate as long as $N$ is held constant. Once we allow for number 
fluctuations, however, we must consider the energy cost $U_0$ of 
removing (adding) a Cooper pair from (to) a layer. Ultimately, this 
cost is related to the electric charge 
neutrality: removing or adding a pair from/to a neutral layer costs Coulomb 
energy.
Thus, we estimate $U_0$ as the electrostatic energy of the charge distribution
corresponding to an extra pair at $\bs$ in layer 1 and a missing pair at
$\bs'$ in layer 2. The missing pair is modeled simply as an effective
positive charge distribution on the otherwise neutral background. We have
\begin{equation}
U_0={1\over 2\epsilon}\int_\bR\int_{\bR'}{\rho(\bR)\rho(\bR')\over |\bR-\bR'|},
\end{equation}
with $\bR=(\br,z)$ a three dimensional vector. $\epsilon$ is the dielectric
constant which reflects the polarizability of the insulating medium between 
the Cu-O layers. Its value is around 10 in YBCO.\cite{tsvetkov}
Taking
\begin{equation}\label{ro}
\rho(\bR)=\rho_1(\br-\bs)\delta(z)-\rho_1(\br-\bs')\delta(z-d),
\end{equation}
where $\rho_1(\br)=(2e)b_d(\br)$ is the planar charge density associated with
a single Cooper pair in PWC, we obtain
\begin{equation}\label{den2}
U_0={(2e)^2\over \epsilon}\int_\br\int_{\br'}b_d(\br)b_d(\br')
\left[ {1\over |\br-\br'|}-{1\over\sqrt{(\br-\br'-\bl)^2+d^2}}\right],
\end{equation}
with $\bl=\bs-\bs'$. For general values of $\bl$ and $d$ the above integral
must be evaluated numerically. However, it turns out that we shall only
require its leading behavior in the limit $d, |\bl| \ll \lambda_d$,
i.e.\ the situation when the distance between the two charge clouds is small compared
to their lateral size given by $\lambda_d$. In this limit we can expand
\begin{equation}\label{den3}
U_0\simeq{(2e)^2\over\lambda_d\epsilon}\left[{d\over 2\lambda_d}+
{\pi\over 8}{\bl^2\over\lambda_d^2}\right].
\end{equation}

The energy correction (\ref{e2}) thus becomes
\begin{equation}\label{e3}
\Delta E^{(2)}=-\int_\bs\int_{\bs'}\left[
{J^2\over U_0+\Delta U}+{J^2\over U_0-\Delta U}\right]
\end{equation}
where we have suppressed various arguments in order to display the structure
of the result. We notice that $\Delta E^{(2)}$ depends on the positions
$\{\br_{i1}\}$ and $\{\br_{i2}\}$ through $\Delta U$. Upon integration
over $\bs, \bs'$ this will produce interactions between vortices
in different layers, as expected. The origin of this interaction lies in the
fact that the energy of the virtual intermediate state depends on the position 
of the extra Cooper pair relative to the pairs already present in that layer. 
Since 
the positions of the former before and after the tunneling event are strongly 
correlated this induces interaction between Cooper pairs in the adjacent
layers. 

We can make the form of the interlayer interaction more explicit in the limit
$|\Delta U|\ll U_0$ by expanding
\begin{equation}\label{e4}
\Delta E^{(2)}\simeq-\int_\bs\int_{\bs'}
{2J^2\over U_0}\left[1
+\left({\Delta U\over U_0}\right)^2 +{\cal O}\left({\Delta U\over U_0}\right)^4
\right].
\end{equation}
The first term is a constant, but the second term, when expanded with 
help of Eq.\ (\ref{den1}), contains the following expression,
\begin{equation}\label{vz}
V_z(\br_{i1}-\br_{j2})=4\int_\bs\int_{\bs'}
V(\br_{i1}-\bs){|J(\bs-\bs')|^2\over U_0^3(\bs-\bs')}V(\bs'-\br_{j2})
\end{equation}
which provides an explicit interaction potential between the pairs located in 
the adjacent layers. Although we derived Eq.\ (\ref{vz}) assuming just two 
layers it is obvious that it generalizes to a multilayer system.

In order to complete our computation of $V_z$ we must determine the form
of the tunneling matrix element $J(\bs-\bs')$. In the direct picture,
modeled by a Lawrence-Doniach type Hamiltonian,\cite{doniach1} Cooper pairs
are assumed to tunnel `straight up' (or down), i.e.\ from a point $\bs$ in 
one layer to the point $\bs$ in the adjacent layer. 
A subtlety in the dual picture arises from the fact that $H_J$ describes
tunneling of {\em dual vortex cores} and not Cooper pairs directly. We recall
that even when the position of the dual vortex is sharply localized at a point 
$\bs$, the
associated Cooper pair charge (and number) density is delocalized over the
length scale $\lambda_d$ around $\bs$ with the probability density given 
by $b_d(\br-\bs)$. Thus, we can think of a Cooper pair as being described 
by a wave function whose envelope varies as $\sqrt{b_d(\br-\bs)}$. As
illustrated in Fig.\ \ref{fig3}.b a straight up tunneling of a Cooper pair 
in general may lead to dual vortex core tunneling with nonzero lateral 
displacement. The associated amplitude to tunnel from point $\bs$ to $\bs'$
will be given by the overlap of the two pair wave functions, 
$\sim\int_\br\sqrt{b_d(\bs-\br)b_d(\br-\bs')}$. This last integral is 
somewhat difficult to evaluate because of the square root, but the resulting
function is clearly close to $b_d(\bs-\bs')$. In the following we thus use
\begin{equation}\label{J}
J(\br)\approx J_0 b_d(\br),
\end{equation}
which turns out to be properly normalized since $\int_\br b_d(\br)=1$. 

An important implication of the above result (\ref{J}) is that tunneling
over lateral distances larger than $\lambda_d$ is exponentially suppressed
and we may indeed use the approximation (\ref{den3}) for $U_0$ when evaluating
$V_z$ from Eq.\ (\ref{vz}). In fact we will use only the first term in 
(\ref{den3}) which gives the leading contribution of the form,
\begin{equation}\label{vz1}
V_z(\br)={4J_0^2e_r^4\over U_0^3}\left[\lambda_d^4\int_\bs\int_{\bs'}
b_d(\br-\bs)b_d(\bs-\bs')^2b_d(\bs')\right],
\end{equation}
where we have used the relation $V(\br)=e_r^2\lambda_d^2 b_d(\br)$ in order
to express everything in terms of $b_d$. It is worth noting that in Eq.\
(\ref{vz1}) the prefactor  
\begin{equation}\label{ej}
E_J={4J_0^2e_r^4\over U_0^3}
\end{equation}
characterizes  the energy scale for 
the interlayer coupling while the expression in the square brackets represents
a dimensionless function of spatial variable $\br$ whose range is set
by $\lambda_d$. The interaction is repulsive.

We could now repeat the calculation of the normal modes for the layered
system. This calculation,\cite{unpublished} however, is lengthy and does not 
yield any new physical insights to the problem at hand. All we need, in fact,
is the result for the speed of sound in the $\hat{z}$ direction. This can 
be readily estimated in analogy with Eq.\ (\ref{vs}) which states that up to
a numerical constant the speed of sound squared equals the energy scale of the
interaction divided by mass. This of course is a well known result for 
ordinary systems of springs and masses; what  Eq.\ (\ref{vs}) confirms is
that this result remains valid even in the case of interactions with finite 
range. We thus obtain
\begin{equation}\label{v_z}
v_z^2\simeq{E_J\over m_v}={4J_0^2e_r^4\over U_0^3m_v}.
\end{equation}

The above is a perturbative result and remains valid only as long as 
$J_0/U_0\ll 1$. With help of Eqs.\ (\ref{den3}) and (\ref{lamc}) this
condition becomes
\begin{equation}\label{pert2}
1\gg{J_0\over U_0}=\left({a \over d}\right)^2{\epsilon\over 16\pi\alpha^2}\left({\lambda_d\over \lambda_c}
\right)^2\simeq 4.1\times 10^2\left({\lambda_d\over \lambda_c}\right)^2,
\end{equation}
where $\alpha=e^2/\hbar c\simeq 1/137$ is the fine structure constant. The
latter appears because the Josephson coupling connects two ``dual worlds'' 
through the real world coupling. Since $\lambda_c$ is typically very large
in cuprates $\sim 10^4-10^5$\AA~ Eq.\ (\ref{pert2}) should be well satisfied
except very close to the transition. 

We now briefly discuss the direct Coulomb interaction, which will turn
out to be negligible in most cases.
If we assume that only the interaction between neighboring layers is important
then standard treatment leads to
acoustic dispersion for the transverse mode along the $\hat{z}$ direction 
with the sound velocity $\sqrt{E_C / m_v}$ where $E_C$ is of 
the order of the Coulomb potential at distance $d$, 
\begin{equation}\label{ec}
E_C = {(2 \nu e)^2 \over \epsilon d} 
\end{equation}
Here
$2\nu e\simeq 2ea^2/\lambda_d^2$ is the fraction of Cooper pair charge that is 
modulated. 
As mentioned above $\nu\approx (1-5)\% $ in the STM experiments.

For the typical values of various parameters listed below Eq.\ (\ref{beta})
and $\lambda_d=20$\AA~ we find $E_J/E_C\approx 30$; the Josephson term dominates. 
It is also important to note that $E_J$ and $E_C$ exhibit very different scaling
with $\lambda_d$. In particular Eq.\ (\ref{ej}) implies that 
$E_J\sim\lambda_d^2$ while $E_C\sim\lambda_d^{-4}$. Thus, even if far away
from the transition the direct Coulomb contribution is significant, as the 
transition is approached and $\lambda_d$ becomes large, the Josephson coupling
between the layers always takes over. In the following we shall thus focus 
exclusively on the latter.


\section{Specific heat and thermal conductivity}\label{section:thermal}

Having found the eigenmodes of the PWC as well as the effective mass of dual 
vortices, we may now combine these results and calculate 
the specific heat and the thermal conductivity associated with the transverse
vibrational modes. As mentioned above the longitudinal modes either have much 
higher speed or are gapped and thus will not contribute at low temperatures.

Let us first estimate the transverse in-plane sound velocity (\ref{vs}) in 
terms of physically meaningful parameters. Combining Eqs.\ (\ref{er}) and 
(\ref{mv})
this becomes
\begin{equation}\label{vs2}
v_s =  c \sqrt{\pi\vartheta\over \ln(\kappa_d)} 
\left({\lambda_{TF} \over \lambda}\right).
\end{equation}
We observe that the dependence on $\lambda_d$ has dropped out, except 
through $\kappa_d$. We expect Eq.\ (\ref{vs2}) to be 
valid only deep in the PWC state; as one approaches the transition point
the dual London approximation ceases to apply.

Thomas-Fermi screening length is normally of the order of
inverse Fermi wave vector $k_F$, or several \AA. The ratio
of the two length scales is thus about $10^{-2}-10^{-3}$. For 
$\kappa_d\simeq 10$ the square root is 0.30. Eq.\ (\ref{vs2}) thus gives
$v_s$ about an order of magnitude smaller than the Fermi velocity in 
cuprates (the latter is $\sim 10^{-2}c$). This makes sense physically since
the PWC vibrations are purely electronic phenomenon, and, thus, on 
dimensional grounds one expects $v_s$ and $v_F$ to be of similar order of
magnitude.

Similarly we can express the interplane sound velocity $v_z$ associated with 
the Josephson coupling. Evaluating Eq.\ (\ref{v_z}) with help of 
 Eqs. (\ref{er}), (\ref{mv}), (\ref{lamc}) and (\ref{den3}) we obtain
\begin{equation}\label{vz2}
v_z =  c{\epsilon^{3/2}\over 8\sqrt{2\ln(\kappa_d)}} 
\left({a_0\over d}\right)^2\left({\lambda_{TF}^2 \over d \lambda}\right)
\left({\lambda_d \over \alpha \lambda_c}\right)^2.
\end{equation}
Thus, $v_z$ grows with $\lambda_d$. According to the criterion (\ref{pert2}) 
the perturbation theory breaks down when $(\lambda_d/\lambda_c\alpha)$ becomes 
of the order of $10$. 
A glance at Eqs.\ (\ref{vs2}) and (\ref{vz2}) confirms that this 
makes sense physically for in this limit $v_z$ approaches $v_s$. The sound
velocity 
becomes isotropic and we may no longer treat the interlayer tunneling as 
a perturbation. 

Repeating the standard calculation of the phonon specific heat for the case
of sound velocity with uniaxial anisotropy leads to
\begin{equation}
C_v = {2 \gamma  k_B^4 \over  \pi^2\hbar^3 v_s^2 v_z} T^3,
\end{equation}
where $\gamma = \int_0^\infty dx x^3/(e^x -1) = \pi^4/15$. 
The associated thermal conductivity in the Cu-O plane becomes
\begin{equation}
\kappa = {1 \over 3} C_v v_s \ell = \beta T^3, 
\end{equation}
with
\begin{equation}\label{eq:beta}
\beta ={2 \gamma  k_B^4 \ell\over  3 \pi^2 \hbar^3 v_s v_z}.
\end{equation}
Combining Eqs.~(\ref{vs2}), (\ref{vz2}) with Eq.~(\ref{eq:beta}) 
we obtain
\begin{equation}\label{beta}
\beta = (0.007 - 3.5)\left({10 {\rm \AA} \over \lambda_d}\right)^2{\rm mW \over K^4 cm},
\end{equation}
where the lower bound obtains if we take $\lambda=1030$\AA~ and 
$\lambda_c=6.2\lambda$ as found for optimally doped YBCO,\cite{esr} while
the upper bound obtains for $\lambda=2020$\AA~ and $\lambda_c=37.7\lambda$,
values relevant to the 56K ``ortho-II'' phase. In both cases we take 
$\lambda_{TF}\simeq a_0=3.8$\AA~,  $d=12$\AA~, $\epsilon=10$ and $\kappa_d=10$.
We assume that the  phonon mean free path $\ell\simeq 0.05$mm, which is the 
geometric average of the width ($\sim 500\mu$m) and the thickness 
($\sim 10\mu$m) of samples used in the experiment.\cite{Taillefer1}

The experimental result\cite{Taillefer1} for the point farthest from
the transition gives  $\beta \approx 0.5$ mW/K$^4$ cm. With the value of 
$\lambda_d=25-50$\AA, estimated from the STM data via Eq.\ (\ref{lambda_d}), 
our result above is 
broadly consistent with the experimental data, although it has to
be noted that it depends strongly on the assumed values of the input
parameters, most notably on the $ab$-plane and $c$-axis penetration depths. 
These are known in YBCO with high accuracy\cite{esr} but they also vary 
strongly with doping and it is not entirely clear which values one should 
adopt in the estimate of $\beta$.
We have argued above that one should take the underlying mean-field,
noninteracting values, which are presumably most accurately approximated
by the values measured near the optimal doping. Within the class of phase 
fluctuation
models considered here, any reduction of superfluid density upon 
underdoping is attributed to an interaction effect beyond mean field theory. 
In reality,
part of the change may be associated with the change of the underlying 
mean-field ground state but it is impossible to make a precise statement
about this. It is also possible that the measured values at optimal doping
already reflect a fair amount of fluctuations and the underlying mean-field
superfluid density should be higher. The range of values displayed in 
Eq.\ (\ref{beta}) is meant to be indicative of these various uncertainties.


\section{Summary and open issues}
We have studied the normal modes of a pair Wigner crystal employing a duality 
transformation from a phase fluctuating superconductor to a 
fictitious dual type-II  superconductor in applied magnetic filed.  
Vortices in the dual superconductor represent Cooper pairs 
in the original model and the vibrational modes of the PWC can be calculated
as magnetophonons of the dual Abrikosov vortex lattice. 
Assuming that that
pinning of the dual vortex lattice to the underlying ionic lattice is 
negligible, as suggested by the fact that PWC is incommensurate in some cases,
the transverse magnetophonon is acoustic and will thus contribute to the
low energy thermal and transport properties of the system. For charged systems
longitudinal modes will be gapped. Our main result is 
the estimate of the sound velocity of the transverse mode which determines
the magnitude of its contribution to the specific heat and thermal conductivity.
The in-plane velocity, $v_s$, is found to be about an order of magnitude lower
than the Fermi velocity $v_F$. Together with our estimate of the 
interplane sound
velocity, $v_z$, which we assumed to be determined primarily by the Josephson
coupling between the neighboring layers, our considerations
yield a $T^3$ contribution to the thermal conductivity with a prefactor
consistent with the recently reported bosonic mode in strongly underdoped 
single crystals of YBCO.\cite{Taillefer1}   

An important length
scale in this problem is the dual penetration depth, $\lambda_d$, which 
has the physical meaning of the size of the Cooper pair in the PWC. STM 
experiments indicate that $\lambda_d$ is much larger than the distance between
the Cooper pairs; the latter are extended objects with strong zero-point 
motion and overlapping wave functions. This is the main reason why duality
is a useful concept in this problem: it provides a convenient tool for the 
description of a 
strongly quantum fluctuating system of Cooper pairs in terms of dual 
vortices that can be treated as point particles. 
The latter are local objects and their physics
can be accurately described in the mean field approximation.

A key assumption, underlying all our preceding considerations and results, 
is that the pair Wigner crystal is essentially decoupled from the ionic 
lattice. As mentioned in the Introduction transverse modes of a PWC pinned
to the ionic lattice would be gapped and thus irrelevant at low temperatures.
The duality
transformation reviewed in Sec.\ II shows that PWC indeed can exist in 
continuum, independently of any underlying crystalline lattice. There is, 
therefore, no logical contradiction implied by the above assumption. In the
context of cuprates one must ask to what extent does the continuum model
reflect the physics of Cooper pairs moving in the copper-oxygen planes. 
The key issue here is whether PWC is commensurate or incommensurate with the 
underlying ionic lattice, since incommensurate PWC {\em cannot} be pinned, 
except
by disorder. Pinning by disorder affects the magnetophonon mean free path 
but in general does not open a gap in the phonon spectrum. 

The problem of commensurability is a difficult one 
to analyze theoretically as it involves the details of PWC energetics, band 
structure and electron-ion interaction. Our argument in favor of the  
incommensurate PWC is therefore largely phenomenological and is based on 
the following three observations. First, the checkerboard patterns in at least 
some experiments\cite{Vershinin,Hashimoto} have been reported to have period
clearly different from 4 ionic lattice constants (the values range between
$4.2-4.7a_0$) implying
incommensurate PWC. The mere existence of such an incommensurate PWC indicates
that coupling to the lattice must be extremely weak even in the case when the PWC
is commensurate. Second, the PWC appears to exist for a relatively wide range 
of dopings. If only commensurate PWC were allowed then one would expect
dramatic changes in PWC structure upon variation of the pair density; in 
particular the unit cell size would vary significantly as the PWC adjusted 
to different doping levels. No such dramatic variations are observed. 
It appears, instead, that PWC structure, where it exists, is largely 
independent of doping. Finally, we found that Cooper pairs in a crystal 
proximate to a superconductor have wavefunctions delocalized over many
lattice constants. We may expect that a potential with ionic lattice 
periodicity should be relatively ineffective in pinning such delocalized 
objects.

An issue which might require further consideration is the dependence of 
the sound velocity of the PWC on the doping level of the system. Experimentally
the strength of the $T^3$ mode goes to zero continuously as the transition
to the superconducting state is approached. In our theory the transition
is marked by the divergence of the dual penetration depth $\lambda_d$.
We found the in-plane sound velocity to be independent of $\lambda_d$. 
This is due 
to the exact cancellation between the strength of intervortex interaction 
$e_r$ and the dual vortex mass $m_v$: both vanish as $\lambda_d\to\infty$.
The $c$-axis sound velocity is proportional to $\lambda_d^2$ which ensures
that the thermal conductivity  of the bosonic modes decreases as the 
transition is approached, in agreement with the experiment. It is important to
emphasize, however, that there is no reason to expect that this result
will remain valid very close to the transition. First, $v_z$ is obtained
in a perturbation theory which is only valid as long as 
$\lambda_d/\lambda_c \ll 1$. Outside of this regime one must treat 
the full 3D system. Second, the dual London approximation also breaks down
near the transition since the vortex cores begin to overlap. 
To address the physics of magnetophonons near the transition one would 
have to treat vortex vibrations in the full 
Ginzburg-Landau theory Eq.\ (\ref{emf1}). Qualitatively we expect that the
amplitude fluctuations will make the intervortex interaction stronger, leading
to increase in $v_s$ and thus reduction of thermal conductivity as the 
transition is approached, in accord with experiment. Detailed calculations,
however, present a daunting challenge and are left for future investigation.

Another open issue is the inclusion of the detailed structure of the PWC. We have 
assumed a simple square Bravais lattice of Cooper pairs. Experiments
\cite{Hanaguri,McElroy,Hashimoto} and detailed theoretical considerations
\cite{Melikyan,balents1} indicate a lattice with square symmetry, but with a 
more
complicated internal structure. Within our approach this could be modeled as a
square vortex lattice with a basis. The vibrational spectrum of such a lattice
will be more complicated but will retain the acoustic mode derived above
which reflects the center of mass motion of the unit cell. 

STM studies also indicate the presence of domain walls and other defects in 
the PWC.
Such defects would scatter PWC phonons and cause a short mean free path
$\ell$. Our estimate, on the other hand, suggests that in YBCO $\ell$ should 
be of the order of the sample size. This implies that the PWC in YBCO is
much more homogeneous than that in BiSCCO and Na-CCOC. 
Given the extreme purity of the YBCO crystals used in the thermal conductivity
measurement\cite{Taillefer1} this is perhaps not surprising. Unfortunately
YBCO is not amenable to high-precision STM studies due to its lack of a natural
cleavage plane. It would be interesting to see if the bosonic mode could be
observed in the thermal conductivity of Na-CCOC. Based on our model we would 
expect it to be much weaker due to much shorter mean free path $\ell$.

{\bf Acknowledgments} The authors are indebted to E. Altman, J.C. Davis, 
T. P. Davis,  N. Doiron-Leyraud, M.P.A. Fisher, 
K. LeHur, L. Taillefer, and Z. Te\v{s}anovi\'c for stimulating discussions
and correspondence. The work has been supported in part by NSERC, CIAR,
A.P.\ Sloan Foundation and the Aspen Center for Physics.


\appendix
\section{Evaluation of $Z_{xy}^{\alpha\beta}$}\label{ap:zxy}
This appendix follows the calculation of Fetter,\cite{Fetter} adjusted to 
the case of the square lattice.  The matrix $Z_{\alpha\beta}^{(2)}$ is a 
symmetric rank-2 tensor. In terms of the vector $\bk$ it can therefore 
be written as
\begin{eqnarray}
Z_{\alpha\beta}^{(2)}(\bk) = A(k^2)\delta_{\alpha\beta} + B(k^2){k_\alpha k_\beta \over k^2}
\end{eqnarray} 
where $A$ and $B$ are scalar functions of $k^2$.  
The trace of $Z^{(2)}$ vanishes to first order in $1/(\lambda_d^2 Q_0^2)$, thus
\begin{eqnarray}
2A(k^2) + B(k^2) = 0.
\end{eqnarray}
We evaluate the off-diagonal elements of the matrix to determine $B(k^2)$.  

Let us define
\begin{equation}
S(\bk) = 2\pi n_v \sum_{Q \neq 0}\left[{(Q_x + k_x)(Q_y + k_y)\over 
(\bQ+\bk)^2} - {Q_xQ_y \over \bQ^2}\right]
\end{equation} 
where $n_v = {1/a^2}$ is the density of vortices.  Ewald summation technique
splits the above sum into two parts, $S =S_d(\bk)+S_r(\bk)$, in such a way
that $S_d$ converges rapidly in real space while $S_r$ converges rapidly in 
reciprocal space. One obtains, for any Bravais lattice,\cite{Fetter} 
\begin{eqnarray}
S_d &=& 2 \sum_j (1-e^{i\bk\cdot \bR_j}){X_j Y_j \over R_j^4}(1+\pi n_v R_j^2)e^{-\pi n_v R_j^2} \nonumber \\
S_r&=&2 \pi n_v \sum_{\bQ \neq 0} {(Q_x + k_x)(Q_y + k_y)\over (\bQ +\bk)^2} e^{-(\bQ+\bk)^2/4 \pi n_v} \nonumber \\ 
 &-&2 \pi n_v {k_x k_y \over k^2}(1-e^{-k^2/4\pi n_v})
\end{eqnarray}
where $\bR_j=(X_j,Y_j)$ is a lattice vector. The above sums can be easily 
evaluated numerically for arbitrary $\bk$.
For small wave vector $|k|$ we may expand $S_d$ and $S_r$ to second order
\begin{eqnarray}
S_d &\approx& 2k_x k_y {\sum_j}' {X_j^2 Y_j^2 \over R_j^4}(1+\pi n_v R_j^2)e^{-\pi n_v R_j^2} \\
S_r &\approx& -{1\over 2}k_x k_y \left[{\sum_Q}'\left( 1-{Q^2\over 8\pi n_v}\right)e^{-Q^2/4\pi n_v}+1\right] \nonumber
\end{eqnarray}
where $\sum'$ denotes a summation that excludes the zero vector.
In order to perform the sums for the square lattice we substitute
\begin{eqnarray}
\bR = a(l,m), \ \ 
\bQ = {2\pi \over a}(l,m)
\end{eqnarray}
and arrive at
\begin{eqnarray}
S_d &=& k_xk_y {\sum_{l,m}}'{2 l^2m^2 \over (l^2+m^2)^2}[1+\pi(l^2+m^2)]e^{-\pi(l^2+m^2)} \nonumber \\
S_r &=& {1\over 2}k_xk_y \bigl({\sum_{l,m}}'[1-{\pi \over 2}(l^2+m^2)]
e^{-\pi(l^2+m^2)}+1\bigr)
\nonumber
\end{eqnarray}
The sums over $m$ and $l$ are rapidly convergent and can be evaluated 
numerically.  This gives the off-diagonal part of $Z^{(2)}$
\begin{equation}
Z^{(2)}_{xy}(\bk) = -0.415{a^2 \over 2 \pi}k_x k_y 
\equiv -\vartheta a^2 k_x k_y,
\end{equation}
with $\vartheta=0.066$.
Thus, $B(k^2)=-\vartheta a^2k^2$.



\begin{thebibliography}{10}
\bibitem{remark1} The dual theory in (3+1)D can be characterized as a
string condensate coupled to tensorial Kalb-Ramond field $B_{\mu\nu}$. The 
latter is related to the electrical 4-current by $j^\alpha_e=2\pi\Phi_0^{-1}
\epsilon^{\alpha\beta\mu\nu}\partial_\beta B_{\mu\nu}$, where $\epsilon$
is the totally antisymmetric tensor.
\end{thebibliography}
\end{document}